\documentclass{article}
\usepackage{arxiv}
\usepackage{multirow}

\usepackage[utf8]{inputenc} 
\usepackage[T1]{fontenc}    
\usepackage{hyperref}       
\usepackage{url}            
\usepackage{booktabs}       
\usepackage{amsfonts}       
\usepackage{nicefrac}       
\usepackage{microtype}      
\usepackage{lipsum}		
\usepackage{graphicx}
\usepackage{doi}

\title{How Developers Choose Debugging Strategies for Challenging Web Application Defects}

\author{ \href{0000-0001-9040-4313}{\hspace{1mm}Maryam Arab}
\\
	School of Information\\
	University Of Michigan\\
	Ann Arbor, MI 48109 \\
	\texttt{maryarab@umich.edu} \\
	\And
    {\hspace{1mm}Jenny T. Liang} \\
	School of Computer Science\\
    Carnegie Mellon University\\
    Pittsburgh, PA 15213 \\
    \And
    {\hspace{1mm}Valentina Hong} \\
	Department of Computer Science\\
    University of Maryland\\
    College Park, MD 20742\\
    \And
    {\hspace{1mm}Thomas D. LaToza} \\
    Department of Computer Science\\
   George Mason University\\
   Fairfax,VA 22030\\
}

\renewcommand{\shorttitle}{HOW DEVELOPERS CHOOSE DEBUGGING STRATEGIES FOR CHALLENGING WEB APPLICATION DEFECTS}

\hypersetup{
pdftitle={Context factors in choosing debugging strategies},
pdfsubject={q-bio.NC, q-bio.QM},
pdfauthor={Maryam Arab, Jenny T. Liang, Valentina Hong, Thomas D. LaToza},
pdfkeywords={Debugging strategies \and Decision-making \and Web application defects \and Contextual factors \and Challenging web application defects},
}

\begin{document}
\maketitle

\begin{abstract}
	Effective debugging is a crucial aspect of software development, demanding a combination of problem-solving skills, expertise, and appropriate tools. Although previous research has studied the effective debugging strategies of expert developers, the specific factors influencing how they choose the debugging strategy, particularly in complex scenarios, remain underexplored. To investigate the contextual factors that add to this complexity and how they affect the expert developers' strategy selection, we conducted two studies. In the first study, we surveyed 35 developers with various expertise to gather a wide range of experiences and identify their recent experience debugging a challenging problem and the problem context factors that add to its complexities. In the second study, we conducted semi-structured interviews with 16 experienced developers to gain deeper insight into the strategic reasoning to choose and replace a strategy to tackle complex debugging tasks. Combining insights from both groups enriched our overall findings, leading to a more robust understanding of debugging strategies at different expertise levels. We found that contextual factors interact in complex ways, and often a combination of factors plays a role in choosing a debugging strategy, potentially evolving throughout the debugging process. In addition, we found that hypothesis making is the baseline for debugging, and experience and familiarity with the code are keys to making the correct decision in both hypothesis generation and choosing relevant strategies. In general, our results show a gap between learning strategies and practicing them effectively in challenging problem contexts, which highlights the need for a more careful design of debugging tools around problem contexts. Our findings also call for the design of educational frameworks that support developers in choosing effective strategies based on the characteristics of their problem context.
\end{abstract}


\section{Introduction}
Effective debugging is a critical skill in software development, which requires substantial problem solving abilities, expertise, and the right tools.  Each step in the debugging process demands careful consideration to achieve success~\cite{Spinellis16EffectiveDebugging, spinellis2003code}. Expert developers bring a vast repertoire of strategies to challenging debugging situations. Using well-defined strategies developed through experience, expert developers navigate the complexities of debugging to effectively resolve complex problems \cite{latoza2010importance, katz1987debugging, Spinellis16EffectiveDebugging, lukey1980understanding, bohme2017bug}. 

Developers employ various strategies such as hypothesis testing \cite{bohme2017bug, robillard2004effective}, following dependencies through execution traces \cite{bohme2017bug, katz1987debugging,romero2007debugging, Spinellis16EffectiveDebugging} or tracing backwards from symptoms to the root cause \cite{
agrawal1993debugging, weiser1984program, ko2004designing, lewis2003debugging, lukey1980understanding, jiang2017programmers, gould1975some, bohme2017bug, engblom2012review, robillard2004effective, SpinellisModernDebugging}. They also simplify complex scenarios into clear and manageable examples to isolate defects more clearly \cite{zeller2002simplifying, SpinellisModernDebugging, evans2022pocket}. Documenting and sharing these strategies in detail can substantially enhance the productivity of less experienced developers \cite{arab2021howtoo, latoza2020explicit}.

However, the true expertise lies not only in the arsenal of strategies, but also in correctly choosing and applying the right strategy to a specific context. Problem solving in software development can often feel like navigating a maze, with many possible dead ends before finding the optimal solution. The ability to select the most suitable strategy for a given problem distinguishes expert developers, making them particularly adept at overcoming complex debugging challenges. Expert developers constantly adapt their strategies to fit different contexts, which is essential for their development and success~\cite{haidry2017identifying}. However, experts' strategies and their cognitive processes are frequently implicit and not widely disseminated. This is particularly crucial in addressing challenging defects, where certain barriers can hinder the efficacy of some strategies or lead developers down unproductive paths. 

However, little is known about how expert developers choose the appropriate debugging strategies based on the contextual characteristics of the problem at hand. Bridging this gap is crucial for several reasons. 
Educating developers on how to select appropriate debugging strategies according to contextual factors can significantly enhance their problem-solving capabilities \cite{latoza2020explicit, ko2019teaching}. 
Additionally, different strategies may require entirely different debugging tools. Tools designed to aid specific strategies (e.g. slicers for backward debugging) may become entirely inapplicable in scenarios requiring a different debugging approach (e.g., delta debugging for defect simplification). A deeper understanding of how developers choose debugging strategies will enable researchers to design more effective tools that are tailored to practical needs.

In this paper, we investigate how expert developers choose debugging strategies in challenging web application problems. Web development presents a unique and challenging subdomain for debugging due to its asynchronous and event-driven nature, which complicates the tracking of causality between event handlers \cite{hong2014detecting, mickens2010mugshot, mutlu2015detecting,zheng2011statically}. Debugging web applications often involves managing diverse environments, devices, and browsers, each potentially exhibiting different behaviors. Moreover, the rapid evolution of web technologies and the widespread use of JavaScript with its event-driven architecture further compound these challenges \cite{ocariza2011javascript, ocariza2013empirical}. These characteristics make web development an ideal first domain to explore in-depth, with lessons that likely apply to other complex and evolving areas of software development.

This paper aims to answer the following research questions:

\begin{enumerate}
    \item[] RQ1: What problem context factors influence the choice of debugging strategy developers use?
    \item[] RQ2: How do defect characteristics affect developers' choice of debugging strategies in web development? 
    \item[] RQ3: How do codebase characteristics affect developers' choice of debugging strategies in web development?
\end{enumerate}
   
To address these questions, we conducted two studies. The first study surveyed 35 developers with various experience to pinpoint the challenging web application debugging experiences they have, the strategies they use, and the factors that make their experience challenging. This varied pool of participants helped surface fundamental obstacles and diverse categories of contextual factors that affect problem-solving approaches. Although the survey provided a broad overview, it lacked the depth to fully understand the decision-making processes of experienced developers. Therefore, we conducted a second study involving semi-structured interviews with 16 expert developers to delve deeper into how they choose debugging strategies, the contextual factors they look at while debugging, and how these factors influence their choice.

The studies revealed that developers consider a combination of factors when choosing a debugging strategy, aiming to select the most effective approach. 
The specifics of the problem's context, mainly the codebase characteristics and defect characteristics, play a significant role in the choice of their strategy. Developers switch between different strategies based on the contextual factors that emerge when tackling understanding the root of a problem. They usually start with hypothesis debugging to make and test their hypotheses and often use other debugging strategies to gain a deeper understanding of the codebase or enhance its readability, extending the utility of these strategies beyond mere debugging. In summary, debugging is a very context-related approach that affects developers' choice of strategy when they gain more knowledge about it. This opens opportunities to design context-specific debugging tools. In addition, educational approaches can teach debugging strategies more effectively by teaching what factors the learner needs to investigate and what tool and strategy fit better the problem.

\section{Related Work}
\label{sec:background}
Our work builds on prior research on the activities and challenges of debugging, the strategies developers use when debugging, and the factors that may influence strategy choice.
\subsection{Debugging Activities and Challenges}
Studies have examined the activities involved in debugging. Developers often follow a simplified scientific method by forming and testing hypotheses \cite{Perscheid2014StudyingTA} and asking and answering questions \cite{sillito2006questions}. Expert developers primarily utilize output data to form hypotheses about defects, while input data plays a less significant role \cite{gould1974exploratory, katz1987debugging, carver1987improving}. 

Developers have identified several challenges that can make debugging defects particularly difficult, especially in cases of ``spaghetti" codebases, inapplicable tools, and faulty assumptions or models that make the stacks grow up ~\cite{eisenstadt1997my}.
Researchers have also examined programming barriers and challenges developers face in web debugging \cite{samudio2022barriers, layman2013debugging}. These include environmental challenges related to tool usability, debugger performance in specific domains (e.g., remote debugging), and limitations of development environments; challenges with multi-threaded applications from the asynchronous nature of web applications; and insufficient information that system logs may provide for effective debugging.

Although previous studies have explored barriers in front-end development \cite{samudio2022barriers, latoza2010importance} and debugging challenges in areas like web services and multi-threaded programming, our research fills the gap by focusing on the contextual factors that contribute to debugging difficulty in web development.

\subsection{Debugging Strategies}
Expert developers use well-defined debugging strategies to effectively identify defects \cite{katz1987debugging, latoza2020explicit, bohme2017bug, vessey1985expertise}. They formulate test theories about potential causes of failures (hypothesis-test) \cite{bohme2017bug, eisenstadt1993tales, gould1975some}, using methods such as code reading and editing \cite{zeller2009programs, layman2013debugging, gould1975some, perscheid2017studying}. Other strategies focus on code navigation \cite{vessey1985expertise, gould1975some}, from an initial state in the program to generate facts and hypotheses to test in forward-reasoning \cite{bohme2017bug, katz1987debugging, romero2007debugging, Spinellis16EffectiveDebugging}, to diagnosing faulty values in the code from the problem symptoms in backward-reasoning \cite{agrawal1993debugging, weiser1984program, ko2004designing, lewis2003debugging, lukey1980understanding, jiang2017programmers, Spinellis16EffectiveDebugging, gould1975some, bohme2017bug, engblom2012review, robillard2004effective, SpinellisModernDebugging}. Although backward-reasoning presents challenges with traditional debuggers that do not support backward debugging execution behavior \cite{agrawal1993debugging}, debugging tools that use slicing techniques \cite{weiser1984program} and omniscient debugging have been proposed to help developers step back through program execution \cite{agrawal1993debugging, ko2004designing, lewis2003debugging}. Knowledge-gathering methodologies and theory-testing approaches play distinct roles in effective debugging \cite{SpinellisModernDebugging}.  
Some work views debugging strategies as occurring in phases, with initial steps before deep debugging, organizational techniques to isolate the root cause, hypothesis generation and investigation, and research of the defect \cite{evans2022pocket, alaboudi2023constitutes}.

Although debugging strategies are documented and the benefit of using strategies in problem solving is studied, no study has examined how expert developers employ these strategies and what contextual factors affect their decision when choosing a strategy to debug a challenging web application. Our study evaluated the important contextual factors and how they contribute to choosing debugging strategies.

\subsection{Strategy Selection}
Several factors have been identified as having a key influence on successful debugging. One critical factor is the familiarity with the codebase and the importance of understanding existing code and comprehending codebases that are designed or written with unfamiliar languages \cite{Ahmadzadeh2005AnAO, gould1975some}. Proficient debuggers demonstrate expertise in knowledge about seven aspects of program understanding \cite{decasse1988review}. Skilled developers who struggle with debugging often lack a strong grasp of program implementation, particularly when dealing with code written by others \cite{decasse1988review}. 
This aligns with Gould's seminal work \cite{gould1975some}, which emphasizes the importance of program comprehension in debugging. Diomidis Spinellis \cite{SpinellisModernDebugging} advocates different approaches based on defect complexity; a '\textit{bottom-up}' approach for defects that have an identifiable cause, while recommending a \textit{'top-down'} approach for more complex defects related to performance, security, and reliability.
The nature of defects also plays a crucial role, with some defects spanning multiple lines of code \cite{lucia2012faults}, linked to past code fixes \cite{gu2010has} or differing in the rationale behind choosing specific fixes \cite{murphy2014design}.

Existing research has primarily focused on the factors that impact successful debugging, such as developer familiarity with the codebase and understanding it, and defect complexity. However, our work is novel in that it investigates the specific contextual factors influencing developers' choice of debugging strategies, particularly for difficult-to-debug problems in web development.

\section{Method}
\label{sec:method}
We conducted two complementary studies to explore how contextual factors affect developers' decision-making when debugging challenging web application defects. 

\subsection{Study 1: Debugging Strategies for Challenging Problems}
\label{sec:study1}
In our first study, we investigate the debugging strategies developers employ when faced with challenging web development problems, how these strategies vary, and the context factors that make certain problems particularly difficult to debug. To address these questions, we conducted a survey study structured into two distinct phases.

\subsubsection{Participants}
To recruit participants for the study, we targeted people with experience in web development. Specifically, we focused on students enrolled in the advanced graduate-level course "User Interface Design and Development" at George Mason University. Historically, this course attracts software developers who are already employed in the field and seek to expand their knowledge. 

We included participants with varying levels of expertise to ensure a diverse sample and observe a more balanced distribution of the frequency and types of challenging problems they faced and the debugging strategies they used to address them. This allowed us to capture broad trends, common misunderstandings, and areas where developers may need additional support or training, which are the basis for the interview in the second study.

An invitation email was sent to all students, offering participation in the study as an alternative to one of their class assignments. To be eligible, participants were required to be at least 18 years old, registered in the course, and familiar with web development technologies, including JavaScript, HTML, and CSS with at least one year of professional working experience. Interested students completed a demographic survey detailing their professional experience as software developers, the number of web applications they had developed, and their years of experience working with frameworks. 

Of the students who responded, 35 were selected for the study. This group included 16 software developers and 16 students, 1 technical lead, 1 software researcher, and 1 math instructor. Their professional work experience varied from 1 to 17 years, with a median of 2 years for the students and 4 years for the others. All participants read and signed a consent form, confirming their agreement to participate in the study by emailing their consent to the first author.

\subsubsection{Procedure} 
The study consisted of two phases. 

In the first phase, participants were asked to describe the most challenging debugging experience they had recently encountered in their web development projects. Specifically, we asked them to detail the nature of the problem, their goal, the technologies they used, and identify the factors that contributed to its complexity and difficulty. This phase aimed to gather a broad spectrum of real-world debugging challenges that developers with diverse expertise levels often face and understand the contextual factors contributing to debugging difficulties. Table \ref{tab:hardproblems} shows the list of all problems and their frequency.

\begin{table}
\centering
\sffamily
\fontsize{8}{8}\selectfont
\caption{Challenging web development experiences in the first study and the number of participants who reported each problem}
\label{tab:hardproblems}
\centering
\begin{tabular}{p{0.85\linewidth}p{0.05\linewidth}}
\toprule
\parbox[c]{\hsize}{\textbf{Problem}} & \parbox[c]{\hsize}{\textbf{\#}}
\\\midrule
\textbf{Inconsistencies}: Inconsistent layout and behavior across browsers or devices; sporadic or non-replicable errors
& 14 \\\midrule
\textbf{Asynchronous Operations}: State changes not reflected in the client side 
& 7\\\midrule

\textbf{General DOM Problems}: Visual defects in UI rendering after HTML changes; interaction defects in HTML elements 
& 6 \\\midrule

\textbf{API}: Debugging API-related defects (e.g., call errors, validating data fields); library set up and documentation checking
& 4 \\\midrule

\textbf{Responsiveness}: Inconsistent behaviour between browsers
& 2 \\\midrule

\textbf{Data}: Understanding data flow between components; data accessing between pages 
& 2 \\\midrule

\textbf{Performance:} concerns for the number of API queries
&1\\\midrule

\textbf{Test}: Writing test for front end
& 1 
\\\bottomrule
\end{tabular}
\label{tab:hard-problems}
\end{table}

In the second phase, participants were assigned to write a comprehensive strategy to address one of the frequently reported challenging problems identified in the first phase. Each participant selected a commonly encountered issue and articulated a step-by-step strategy that they would use to debug that problem. This phase allowed us to capture detailed strategies and identify commonalities and variations in the approaches used by different developers.

\subsubsection{Analysis}

Our analysis focused on understanding what strategies developers use to solve challenging web development daily problems, and what problem context factors make the debugging process challenging. To analyze the strategies' content and the survey responses, we followed recent best practices guidelines by  Hammer and Berland~\cite{hammer2014confusing}, which consider the disagreements as interpretation variance.

To characterize how the authors chose to express the strategies, our main focus was to understand the goal of the strategy. Therefore, we evaluated the strategies as a one-piece data record, rather than breaking down the strategy content to statements. For each strategy, the qualitative coding focused on features and subgoals of single or multiple statements of the strategy. 

To analyze the responses of the survey, we created a separate document for each question and its responses for qualitative analysis. We created three data sets with the participants' responses on the nature of the problem, their goals, and the factors contributing to its complexity and difficulty for qualitative coding. In the first round of qualitative coding, the first two authors separately identified the codes with a brief description. The two authors then individually labeled each response with zero or more codes. To aggregate these codes, for each document, the authors first compared the individually generated codes and identified codes with similar definitions to be merged and added to the code book. Then they compared the labeled codes, discussed instances of disagreements, and reached agreements by adding or removing the code from the code book. The disagreements mainly stemmed from the scope of the generated codes or a code from one author included multiple codes from the other author. So, the agreements were resolved by agreeing on the scope they wanted to report the code in the results. The authors then applied pattern coding to the final code book, which grouped the codes into several broader categories. Using the final code book, the authors then coded the data in the second round and resolved any conflicts.

\begin{figure}
\centering
\includegraphics[trim=42 515 62 10, clip, width=1\linewidth, keepaspectratio]{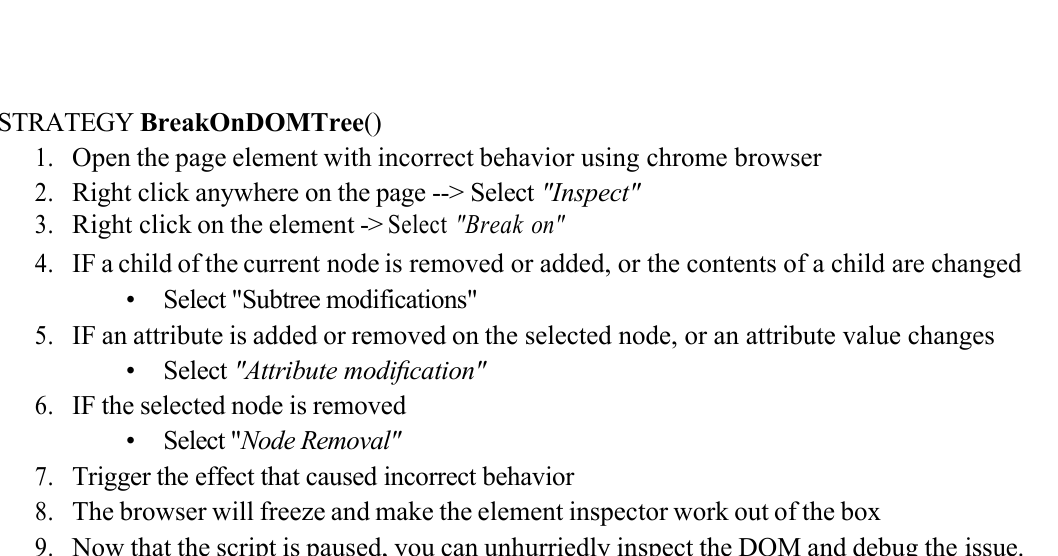}
\caption{Example of explicit programming strategies for debugging the source of defects related to DOM element}
\label{figure:DomDebugging}
\end{figure}
\subsubsection{Results} 
\label{sec:results-firstStudy}

\begin{table*}[]
\centering
\sffamily
\fontsize{8}{8}\selectfont
\caption{The factors contribute to the difficulties of debugging in the first study}
\label{tab:study1-factors}
 \begin{tabular}{p{0.42\textwidth}p{0.52\textwidth}}
\toprule
\parbox[c]{\hsize}{\textbf{Code \& Description}} & \parbox[c]{\hsize}{\textbf{Representative Quote}} \\\midrule
\parbox[c]{\hsize}{\emph{Client-side}---The defect is visual and related to the front-end and CSS files}& 
\parbox[c]{\hsize}{\textit{``Debugging elements using CSS is typically difficult (...) It is difficult to see what changed in the overall layout.''}(P8)}
\\\midrule
\parbox[c]{\hsize}{\emph{Performance}---The number of data retrieval from APIs are limited}& 
\parbox[c]{\hsize}{\textit{``There was performance concern for calling the API too many times to query data from the database. ''}(P17)}
\\\midrule
\parbox[c]{\hsize}{\emph{External Libraries}---The lack of information on the defects related to external resources}&
\parbox[c]{\hsize}{\emph{``API timing out without sending a timeout error back through the AJAX call, and so not all of the results were being received from the API.''} (P16)}
\\\midrule
\parbox[c]{\hsize}{\emph{Deprecated}---Deprecated old/unfamiliar codebase with no documentation}& 
\parbox[c]{\hsize}{\emph{``Lack of documentation requires more time to understand or remember what is occurring where.''}(P5)}
\\\midrule
\parbox[c]{\hsize}{\emph{Familiarity}---lack of knowledge about a language, technology, framework, etc.}&{\parbox[c]{\hsize}{\textit{``Not much experience in front end development, along with little experience in angular made this problem harder than it should have been.''}(P6)}}
\\\midrule
\parbox[c]{\hsize}{\emph{Environment}---typical debugging approach not applicable in specific environments} &
{\parbox[c]{\hsize}{\textit{``Typically console log just work, but this does not work in chrome extension.''}(P3)}}
\\\midrule
\parbox[c]{\hsize}{\emph{Sporadic}---The intermittent and non-reproducible defect} &
{\parbox[c]{\hsize}{\textit{``The problem was intermittent-I would get the correct answer 1 every 2 times at some tries, and the correct answer all 10 times at other tries.'' }(P16)}}
\\\midrule
\parbox[c]{\hsize}{\emph{Compatibility}---Inconsistent behavior across various browsers, devices, and environements}&
{\parbox[c]{\hsize}{\textit{``This problem is hard to solve, as one can never know which device the user might probably use to access the web-application.''}(P29)}}
\\\bottomrule
\end{tabular}
\end{table*}

In general, participants authored three categories of debugging strategies:

\emph{Backward-reasoning} guides developers on how to pause code execution to diagnose possible causes of defects. It involves tracing the program execution flow backward from the point of failure to identify the root cause (similar to backward-reasoning in Table \ref{tab:LiStrategies})

\emph{DOM-debugging} focuses on various scenarios of misbehavior in DOM elements, such as: (1) missing element behavior,
(2) addition or removal of child elements from the currently selected node,
(3) changes in the contents of a child element,
(4) Addition or removal of attributes on the currently selected node, (5) changes in attribute values, and
(6) removal of the currently selected node. For example, the strategy illustrated in Figure \ref{figure:DomDebugging} demonstrates how to use the browser developer tool to stop DOM rendering after specific elements exhibit misbehavior. This strategy provides step-by-step guidance on how to take advantage of different developer tool features to diagnose and resolve DOM-related issues. This strategy is similar to the simplification strategy in Table \ref{tab:LiStrategies}.

\emph{Devtool-debugging} outlines how to use various panels within the browser developer tool, such as "Watch", "Call Stack", and "Scope", to diagnose and address defects. Participants described how to make inline changes to the code using these panels to identify the underlying source of the defect. This strategy is also a sub-strategy of the backward-reasoning strategy in Table \ref{tab:LiStrategies}.

The results of the analysis also revealed a list of factors that contribute to the difficulties of debugging experiences the participants had and are listed in Table \ref{tab:study1-factors}. This list is refined and used as a preliminary source in the design of the second study. Hence, we will describe them in more detail in Section \ref{sec:results-secondStudy} to reduce duplicates.

\subsection{Study 2: Factors Influencing Debugging Strategy Selection}
\label{sec:study2}

Although the first study provided an overview of the factors that contribute to a challenging debugging experience, it lacked the depth needed to fully understand the decision-making processes of highly experienced developers when dealing with complex debugging issues. To this end, we conducted semi-structured interviews with highly experienced web developers.

To ensure that our initial list of debugging strategies and the factors influencing their efficacy was comprehensive, we conducted a \textit{data enrichment} process by reviewing the related literature that documented debugging strategies and the factors affecting their effectiveness. 

\subsubsection{Data Enrichment}
\label{sec:data-enrichment}
The results of the first study informed the design of the second study. To enrich our initial list of debugging strategies, we incorporated insights from existing research on effective debugging practices. The primary sources of debugging strategies, including the strategies of the first study and the strategies from related work \cite{evans2022pocket, Ahmadzadeh2005AnAO, lukey1980understanding, jiang2017programmers, Spinellis16EffectiveDebugging, gould1975some, bohme2017bug} were generalized and are described in Table \ref{tab:LiStrategies}.

Furthermore, we compile a list of contextual factors that influence the selection of developers' strategies based on prior research \cite{Spinellis16EffectiveDebugging, SpinellisModernDebugging, evans2022pocket}. From this process, we were only able to add a couple of more factors to the results of the first study: clear error message, testability, and data related, which confirms the gap in the literature about the contextual factors affecting developers' choosing an effective strategy. These results are shown and cited in Tables \ref{tab:defect-factors} - \ref{tab:other-factors}. These factors were categorized into four main groups: (1) codebase characteristics, (2) defect characteristics, (3) organizational context, and (4) tool availability and usability.

To ensure comprehensive coverage of a diverse set of web problem scenarios, we gathered a list of ten common web development issues reported in prior studies \cite{Ahmadzadeh2005AnAO, katz1987debugging, layman2013debugging}. These scenarios included
production defects, concurrency issues, multi-threaded race conditions, external component defects, large or unfamiliar codebases, unsupported libraries, poorly documented legacy code, compiler-optimized code, multi-instance event handling, and memory exhaustion defects. 

This enriched data set--including the debugging strategies, the contextual factors, and the challenging debugging scenarios, which combined research insights and real-world scenarios-- served as a foundation for our detailed exploration of the factors influencing debugging strategy selection during the semi-structured interviews in the second study.

\begin{table*}
\caption{Six debugging strategies identified in previous research. Each strategy outlines a sequential approach for troubleshooting and resolving defects. The complete definition of the strategies can be found in the supplemental materials.}
\centering
\sffamily
\fontsize{8}{9}\selectfont
\begin{tabular}{p{0.22\textwidth}p{0.58\textwidth}p{0.1\textwidth}}
\toprule
\parbox[c]{\hsize}{\textbf{Strategy}} &
\parbox[c]{\hsize}{\textbf{Description}} &
\parbox[c]{\hsize}{\textbf{Ref}} 
\\\midrule
\parbox[c]{\hsize}{\textit{1. Hypothesis-test}} &
\parbox[c]{\hsize}{Developing hypotheses about the cause of the defect and testing these hypotheses by gathering evidence in the code or runtime behavior.
} &
\parbox[c]{\hsize}{ \cite{bohme2017bug, robillard2004effective, Spinellis16EffectiveDebugging}}
\\\midrule

\parbox[c]{\hsize}{\textit{2. Backward-reasoning}} &
\parbox[c]{\hsize}{Identify solutions or diagnoses by tracing the error's manifestations back through the code execution path to uncover the underlying possible causes.
} &
\parbox[c]{\hsize}{\cite{
agrawal1993debugging, weiser1984program, ko2004designing, lewis2003debugging, lukey1980understanding, jiang2017programmers, Spinellis16EffectiveDebugging, gould1975some, bohme2017bug, engblom2012review, robillard2004effective, SpinellisModernDebugging}} 
\\\midrule

\parbox[c]{\hsize}{\textit{3. Forward-reasoning}} &
\parbox[c]{\hsize}{Starting with an initial state or known facts. move forward logically from the initial state, applying rules, operations, or statements to generate new facts or states, systematically examining each step towards the goal or the discovery of a defect root.  
} &
\parbox[c]{\hsize}{ \cite{bohme2017bug, katz1987debugging,romero2007debugging, Spinellis16EffectiveDebugging}}
\\\midrule

\parbox[c]{\hsize}{\textit{4. Simplification}} &
\parbox[c]{\hsize}{Breaking down the problem into smaller, more manageable parts, removing unnecessary details, and focusing on the core aspects that are crucial for finding a solution.} &
\parbox[c]{\hsize}{ \cite{zeller2002simplifying, SpinellisModernDebugging, evans2022pocket}}
\\\midrule

\parbox[c]{\hsize}{\textit{5. Error-message}} &
\parbox[c]{\hsize}{
Understanding the content and meaning of error messages, which often include error codes, descriptions, and context about where and why the error occurred. followed by reading documentation, online resources, forums, and knowledge bases to look up error.
} &
\parbox[c]{\hsize}{\cite{SpinellisModernDebugging, spinellis2006debuggers}}
\\\midrule

\parbox[c]{\hsize}{\textit{6. Binary-search}}&
\parbox[c]{\hsize}{Repeatedly dividing the codebase or input space into smaller sections and testing each section to isolate the problematic area.} &
\parbox[c]{\hsize}{\cite{SpinellisModernDebugging, evans2022pocket}}
\\\midrule
\end{tabular}

\label{tab:LiStrategies}
\end{table*}

\begin{table*}
\caption{Additional debugging strategies  described by expert developers in Study Two}
\centering
\sffamily
\fontsize{8}{8}\selectfont
\begin{tabular}{p{0.22\textwidth}p{0.58\textwidth}p{0.1\textwidth}}
\toprule
\parbox[c]{\hsize}{\textbf{Strategy}} &
\parbox[c]{\hsize}{\textbf{Description}}  &
\parbox[c]{\hsize}{\textbf{Participant}}
\\\midrule
\parbox[c]{\hsize}{\emph{7. System-level}} &
\parbox[c]{\hsize}{If the defect is user-specific, then verify the user’s role in the database and its restrictions.

Examine the Content Management System (CMS) for the user's role permission settings.}&
\parbox[c]{\hsize}{ [P6]}
\\\midrule

\parbox[c]{\hsize}{\emph{8. External-resources}}&
\parbox[c]{\hsize}{Consult an expert, Google, Large Language Models.}&
\parbox[c]{\hsize}{[P6]}
\\\midrule

\parbox[c]{\hsize}{\emph{9. Historical-analysis}}&
\parbox[c]{\hsize}{Modified version of binary search using ``git-bisect'' command: Isolate the exact commit that introduced a bug by systematically halving the codebase history and testing each version.} &
\parbox[c]{\hsize}{[P7,10]}
\\\midrule
\end{tabular}

\label{tab:strategies-study2}

\end{table*}

\subsubsection{Participants}
In the second study, we recruited expert web developers with: (1) at least eight years of professional experience in web development and (2) a minimum of three years of experience mentoring teams. Our past studies have shown that the experience of mentoring correlates with the ability to write effective strategies \cite{latoza2020explicit, arab2022exploratory}, as mentors often work with learners of varying levels of knowledge and understanding.

Recruitment methods included social networks, email outreach, snowball sampling, and web development Meetups. The recruitment post detailed the time commitment and included a demographic survey for eligibility screening.
Fifty-four developers responded, with 31 meeting the eligibility criteria. We interviewed 16 participants, achieving data saturation. The participants included senior full-stack developers, software development managers with decent experience in web development, and senior web developers with 8 to 38 years of experience (median: 12 years, average: 15 years), contributing to 5 to 70 software projects each.

\subsubsection{Procedure}
We conducted 90-minute semi-structured interviews, with three paper authors participating, one for taking notes and the other two for conducting the interviews and asking follow-up questions. Interviews were audio-recorded and transcribed for analysis. 

The interviews consisted of two phases. The first phase lasted 30 minutes and focused on two of the most frequently reported challenges identified in the first study: inconsistent behavior and asynchronous operation defects (the first two problems listed in Table \ref{tab:hard-problems}). The participants were provided with examples, a list of common debugging strategies (Table \ref{tab:LiStrategies}), and the four categories of contextual factors generated in Section \ref{sec:data-enrichment}. Participants were asked to choose strategies and describe the factors considered in choosing a debugging strategy,
how each factor influenced their choice, and
which strategies would not be effective for their specific case.

The second phase lasted 60 minutes and focused on participants' past debugging experiences to provide real-world scenarios, making it easier for participants to articulate the reasoning behind their decisions and the choices they considered. Participants selected three recent problem scenarios they had encountered from a list of ten common web development problems \cite{Ahmadzadeh2005AnAO, katz1987debugging, layman2013debugging}, including production defects, concurrency issues, multithreaded race conditions, external component bugs, large or unfamiliar codebases, unsupported libraries, poorly documented legacy code, compiler-optimized code, multi-instance event handling, memory exhaustion, and general DOM defects. The participants described their previous experiences and the debugging strategies they used, the factors they considered when choosing or abandoning a strategy, and how each factor affected their choice, providing valuable information about their decision-making process. 

\subsubsection{Analysis}
Our analysis focused on answering three research questions.

\begin{enumerate}
    \item What contextual factors influence developers' choices of debugging strategies?
    \item How do defect characteristics contribute to the developers' choice of strategies?
    \item How do codebase characteristics influence developers' debugging strategy selection?
\end{enumerate}

In the second study, we transcribed and segmented all interviews into manageable units that balanced conciseness with the need for contextual richness. This ensured that each segment was clear and comprehensible, without being overly brief or too complex. To analyze the data, we followed the guidelines established by Hammer and Berland \cite{hammer2014confusing}.
We conducted two rounds of coding and one round of thematic analysis to extract codes related to contextual factors and the causal relationship between factors and debugging strategies. 

First, to extract contextual factors, three authors used the preliminary context factor codebook (as described in Section \ref{sec:data-enrichment}) to perform open coding in interview transcripts. This process involved identifying existing and new codes linked to contextual factors. During author meetings, discussions of code definitions helped ensure uniformity by merging similar definitions and assigning distinct code titles.

Second, to understand the relationship between contextual factors and strategy choices, we performed the second round of coding and employed \textit{'causal coding'} \cite{patel2014guide} to locate, extract or infer causal beliefs from the data and identify relationships between context factors and debugging strategies.

\emph{Codebook Refinement and Thematic Analysis---} After coding, two authors reviewed and refined the codebook. They grouped similar factors into higher-level categories. The finalized codebook was then used to categorize the contextual factors and the reasons for choosing specific debugging strategies. This iterative process facilitated the identification of recurring themes, which are presented in the next section.

\section{Results}
\label{sec:results-secondStudy}

The participants described the alternative debugging strategies they could choose that were not already identified in the debugging strategies we found in previous work (Table \ref{tab:strategies-study2}). We identified six categories of factors that influence developers' choice of debugging strategies: (1) codebase characteristics, (2) defect characteristics, (3) organizational context, (4) tool availability and usability, (5) individual developer traits, and (6) project requirements. Each category contains several factors described in Tables~\ref{tab:defect-factors} to \ref{tab:other-factors}. The factors that form the first study and related works are marked with an asterisk (*).
We describe each group of factors in the following sections.

\begin{figure*}[ht!]
\centering
\includegraphics[trim=0 185 110 90, clip, width=1\linewidth, keepaspectratio]{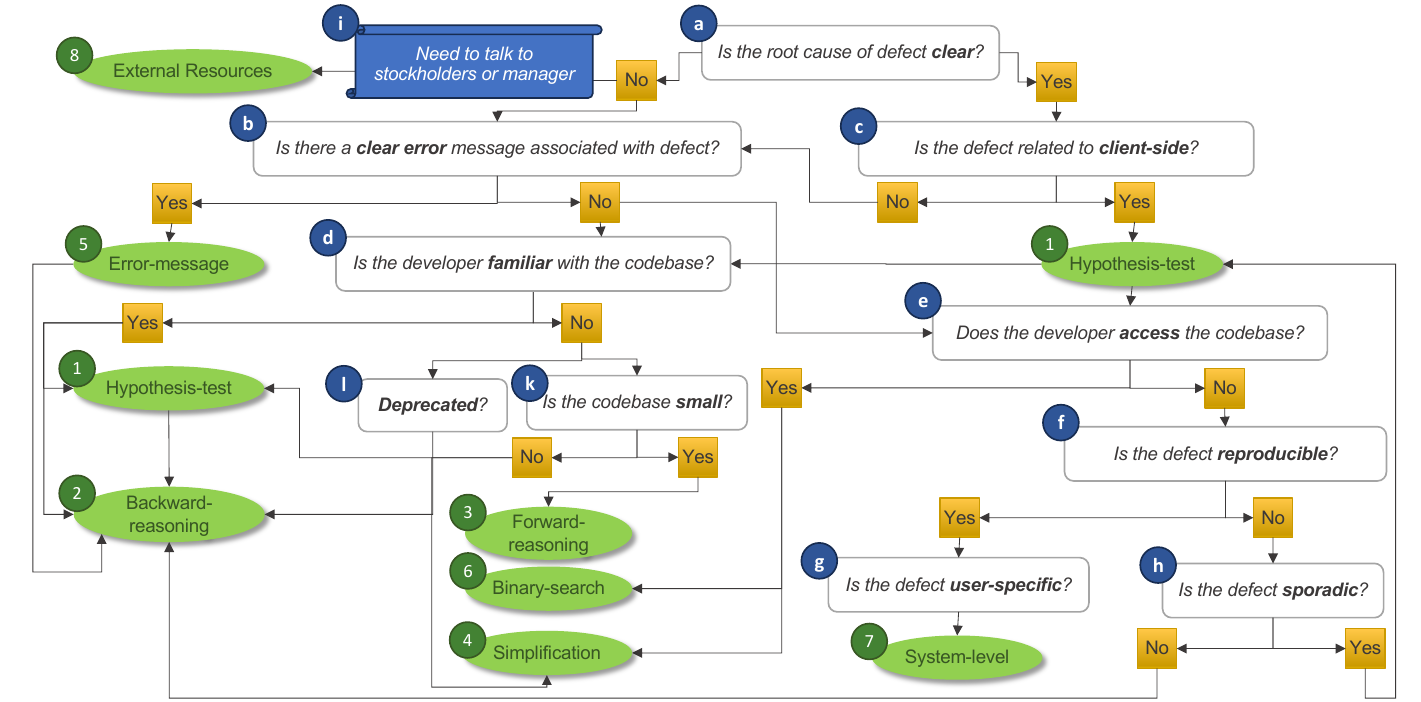} 
\caption{Model of developer decision-making in web debugging strategy selection. Italic text boxes represent factors influencing strategy selection (defect and codebase characteristics in Tables \ref{tab:codebase-factors}, and \ref{tab:defect-factors}).
Green circles represent debugging strategies.}
\label{fig:defect-factors-strategies}
\end{figure*}

\subsection{Defect Characteristics}
\label{sec:contextfacotrs:defect}

\begin{table*}[]
\centering
\sffamily
\fontsize{8}{8}\selectfont
\caption{The context factors related to \textbf{defect characteristics} developers consider when selecting a debugging strategy. 
Codes extracted from our initial previous works review are started (*) and the ones we describe in more details are \underline{underlined}.
}
\label{tab:defect-factors}
 \begin{tabular}{p{0.04\textwidth}p{0.2\textwidth}p{0.65\textwidth}}
\toprule
\parbox[c]{\hsize}{\textbf{Factor }} & \parbox[c]{\hsize}{\textbf{Categories}}
 & \parbox[c]{\hsize}{\textbf{Description }}\\\midrule
 
\parbox[C]{2mm}{\multirow{5}{*}[-6ex]{\rotatebox[origin=c]{90}{\textit{\underline{Clarity}}}}} &
*Clear-error-message & Presence of clear error messages in the execution stack \cite{spinellis2006debuggers, evans2022pocket, layman2013debugging, gould1975some}
\\&
Complexity & Code complexity that complicates isolating the code segments related to the specific behavior causing the defect
\\&
Intended Feature & Whether the reported issue is a genuine defect that needs fixing or an intended feature/behavior. 
\\&
Environment & Whether the defect is associated with an incomplete/failed deployment in the production or an issue during development environment
\\&
& Whether the defect is visual and related to the front-end or the back-end functionalities and overall state of the program's execution.
\\\midrule

\parbox[c]{5mm}{\multirow{5}{*}{\rotatebox[origin=c]{90}{\textit{\underline{Reproducibility}}}}} &

*Sporadic & Defects that occur intermittently, making them difficult to reproduce consistently \cite{layman2013debugging, evans2022pocket}
\\& 
User-specific & Defect visible only to a specific user, not replicate in simulated user’s context or permissions
\\&
*Compatibility & Different defect replication across various browsers, devices, or versions.
  Specific environmental configurations affect the replicability of the defect.
\\&
*Performance & Application crashes, freezes, slow rendering, high memory usage, or excessive requests \cite{layman2013debugging, evans2022pocket}
\\\midrule

\parbox[c]{1cm}{\multirow{5}{*}{\rotatebox[origin=c]{90}{\textit{Root}}}}&

Network-related & Whether the defect is related to API calls, timeouts, or failures in network services.
\\&
*Data-related & Defects involving data aspects, such as size, format, or response status\cite{SpinellisDifferntialDebugging, gould1975some}.
\\&
Configuration & Defects associated with system or user settings and permissions
\\&
Hardware-related & Defects related to the hardware or machine used for hosting or running the application
\\\bottomrule

\end{tabular}
\end{table*}

Our study results showed that expert developers consider specific attributes and properties of defects to determine the most effective approach to identify, isolate, and resolve them. By considering defect characteristics factors, developers can select strategies tailored to the unique challenges of each defect, ultimately improving debugging effectiveness and software quality. These characteristics are described in Table \ref{tab:defect-factors}.  We discuss the \underline{underlined} factors and the categories that developers described would affect their strategy choices in the next sections.

\subsubsection{Clarity}
Developers encounter several factors that can cloud their initial assessment of the root cause of the defect. The clarity of a defect significantly influences the choice of debugging strategies (Fig~\ref{fig:defect-factors-strategies}-a). Clarity encompasses factors such as error message transparency, code complexity, intended versus actual behavior, and the environment (production versus development). Clear defects allow for direct and efficient debugging methods, whereas ambiguous or complex defects require more investigative and strategic approaches.

\paragraph{Clear error messages} 
Clear error messages on the execution stack provide direct pointers to problematic code, enabling straightforward debugging techniques. When error messages are clear (error messages in the error stack or exceptions thrown by the code), developers can quickly identify and resolve issues through direct methods such as immediate code inspection or simple debugging tools (Fig.~\ref{fig:defect-factors-strategies}-b). They leverage \textit{error-message} debugging to trace the problem execution backward. Without clear messages, developers form \textit{hypotheses} based on symptoms and strategically insert temporary code snippets to track execution flow, helping focus on the problematic area before applying \textit{backward-reasoning} to pinpoint the exact location. This approach is most effective when the developer is familiar with the codebase and can make an informed guess based on their past modifications. Once the focus area is identified, developers typically switch to \textit{backward-reasoning} to pinpoint the exact location of the problem.

\paragraph{Complexity} Complex code can obscure connections between symptoms and root causes, complicating isolation and understanding. When symptoms and root causes are not easily connected, developers use \textit{binary-search} techniques or \textit{simplification} strategies to narrow down the specific behaviors causing the defect. These strategies help decompose complex code into more manageable parts.

\paragraph{Intended Feature}
Developers may be uncertain if the observed behavior is a defect or intentional. For example, one participant described a scenario where a delay in updating a user’s account balance might or might not have been intentional. Before initiating debugging, developers can consult the project documentation to clarify requirements or discuss behavior with stakeholders (Fig.~\ref{fig:defect-factors-strategies}-i, 8). This collaborative approach helps reduce confusion for both developers and end users. Stakeholders can help clarify if a user interface design causes confusion. In these scenarios, adding a new feature to clarify the interface may be required instead of debugging.

In other cases, developers often check existing test cases or documentation related to the observed behavior to determine if the behavior is a defect or intended. This ensures time is not wasted fixing non-issues and clarifies whether debugging or a feature enhancement is needed. In such cases, the availability of test cases is a crucial factor in troubleshooting. A representative participant said:

\begin{quote}
    \textit{
    ``Is this a known issue? The first thing is to make sure. Hopefully, there are tests related or something that could help verify this behavior is intended.''
    } (P8)
\end{quote} 

\paragraph{Environment} The defect environment (production or development) helps dictate the debugging strategy. In production, debugging is often constrained by limited access and the need to avoid disrupting users, leading to reliance on logs and indirect methods. In development, developers have the option to use interactive debugging tools and \textit{forward-debugging} techniques. For instance, for post-deployment defects, developers consider recent changes and deployment status. In development, they may set breakpoints and actively debug execution flow \textit{(backward-reasoning)}, while in production, they rely on logging and monitoring tools to diagnose issues with minimal disruption.

Defects related to client or server-side issues also require different strategies. For visual defects or user interface issues that occur in the user's browser, developers often start by forming a \textit{hypothesis} about the root cause. When they have code access (Fig~\ref{fig:defect-factors-strategies}-e), developers may use \textit{simplification} or \textit{binary-search} debugging. These strategies fit client-side issues because removing visual elements does not affect core functionality. As one participant noted:
\begin{quote}
    \textit{``It is safer to delete portions of UI code and put them back without breaking things.''} (P1)
\end{quote}

However, these strategies are ineffective for defects visible only in the deployed production version, where code access is limited.  Removing code in production is typically not possible due to its immediate negative consequences for users.

Participants also reported limitations in debugging client-side issues. Setting breakpoints directly on HTML is not possible, making the \textit{ backward-reasoning} strategy less effective. Relying solely on \textit{error-messages} is not helpful as HTML does not generate errors that halt the execution stack.

In contrast, \textit{simplification} or \textit{binary-search} debugging may not be ideal for server-side defects. Removing code on the server can introduce unintended consequences, potentially creating new defects. If the debugger knows which components, libraries, or functionalities are responsible for the defect, they may rely on \textit{backward-reasoning}. 
When there is no clear cause (Fig~\ref{fig:defect-factors-strategies}-b), developers add log statements to narrow down the problem. However, the effectiveness of this strategy depends on familiarity (Fig~\ref{fig:defect-factors-strategies}-d) with the codebase and its size (Fig~\ref{fig:defect-factors-strategies}-k). In large, unfamiliar codebases, adding many log statements may add confusion. In such cases, developers may revert to \textit{binary-search} or \textit{simplification}. We will discuss the effect of familiarity on strategy selection in more detail in Section \ref{sec:contextfacotrs:codebase}.

\subsubsection{Reproducibility} 
The reproducibility of a defect significantly influences the ease with which a developer can pinpoint its root cause. 

\paragraph{Sporadic}
Sporadic defects appear intermittently and inconsistently, making them particularly challenging to debug. These defects often require extensive logging, user reports, and specific actions to trigger the issue. The ease of reproducing a defect affects the choices of the debugging strategy. Figures \ref{fig:defect-factors-strategies}-f, h show how inconsistent behavior of a defect in different runs can affect the choice of strategy.

The most effective strategy for tackling inconsistent behaviors is \textit{hypothesis-test} debugging. This approach requires two key elements: codebase familiarity and observational skills. A strong understanding of the codebase allows developers to formulate educated guesses about where the issue might lie. Close attention to program behavior during successful and failed runs is important to identify subtle differences in program state, variable values, or execution flow. A representative participant said the following. 

\begin{quote}
    \textit{``It [hypothesis-testing] is a little bit easier to do if the bug is nondeterministic because even if you do not see the bug on a particular run you might still be able to prove or disprove that hypothesis.''} (P12)
\end{quote}

\textit{Binary-search} debugging generally does not work for inconsistent defects, as removing code to isolate the defect does not guarantee problem resolution due to the sporadic nature of the issue. However, in cases where a specific sequence of inputs consistently triggers the bug, \textit{binary-search} might help narrow down the affected code section responsible for handling that particular sequence:

\begin{quote}
    \textit{``A particular sequence of inputs that got to the bad state, one could think about trying to use binary search debugging''}(P12)
\end{quote}

Inconsistent defects pose challenges for \textit{backward-reasoning}, especially if the defect is timing dependent. Breakpoints or log statements can alter the timing, causing the defect to disappear. However, a representative participant noted that intermittent behavior is usually linked to the server-side rather than the visual part of the application, which is directly linked to \textit{backward-reasoning} debugging:

\begin{quote}
    \textit{``HTML is not executing and it is just rendered, so the inconsistent behavior is related to JS that is executing the code.''} (P3)
\end{quote}

\paragraph{User-specific} These defects occur only for specific users due to unique configurations or data specific to their environment. Debugging strategies involve simulating or replicating the user environment through impersonation or analyzing user-specific logs. For user-specific defects (Fig~\ref{fig:defect-factors-strategies}-g), developers may use impersonation techniques and examine user-specific logs to replicate and diagnose the issue. Also, some participants employed \textit{system-level} strategies such as logging in as the user or checking user settings.







\subsection{Codebase Characteristics}
\label{sec:contextfacotrs:codebase}

\begin{table*}[htbp]
\centering
\sffamily
\fontsize{8}{8}\selectfont
\caption{The context factors related to \textbf{code-base characteristics} developers consider when selecting a debugging strategy. 
Codes extracted from our initial previous works review are started (*) and the ones we describe in more details are \underline{underlined}.
}

\label{tab:codebase-factors}
 \begin{tabular}{p{0.1\textwidth}p{0.16\textwidth}p{0.67\textwidth}}
\toprule
\parbox[c]{\hsize}{\textbf{Factor }} & \parbox[c]{\hsize}{\textbf{Categories}}
 & \parbox[c]{\hsize}{\textbf{Description }}\\\midrule

\multirow{3}{*}{\parbox[c]{\hsize}{\emph{\underline{Familiarity}}}} &
*Hands-on Experience & Whether the developer is working on their own code and their familiarity with the programming language and codebase \cite{Ahmadzadeh2005AnAO, decasse1988review, gould1975some}
\\&
Past-exposure & Whether the developer encountered similar defects in the past
\\\midrule

\multirow{5}{*}{\parbox[c]{\hsize}{\emph{\underline{Access}}}} &
\parbox[c]{\hsize}{Source code} & Access to view, edit, and understand the source code, whether on a local machine or a server 
\\&
\parbox[c]{\hsize}{Resources} & Availability of the environments, resources, and devices where the error occurs, such as different browsers, open-source API source code, production environment, operation systems, or hardware configurations 
\\&

\parbox[c]{\hsize}{Level} & The extent of access (full/partial) to various systems and components involved in the application, including the codebase, configurations, or deployment environments 
\\\midrule

\multirow{1}{*}{\parbox[c]{\hsize}{\emph{Technical}}} &
Tech Stack & The frameworks and other technologies used and their unique features, tools, and conventions
\\\midrule

\parbox[c]{\hsize}{\emph{Testability}}&
*Test Case Coverage  & 
The presence and accessibility of resources for testing: test cases, test data, test environments, and tools. The extent to which the existing test cases cover the codebase. 
\cite{layman2013debugging, bohme2017bug, SpinellisModernDebugging}
\\\midrule

\multirow{6}{*}{\parbox[c]{\hsize}{\emph{Complexity}}} &
Modularity & The organization of code into well-defined, manageable, and interchangeable elements 
\\&
Dependency & Inter-dependencies between elements within the code which may require understanding other elements
\\&
Coding Styles & Variability in code styles that impacts readability and maintainability
\\\midrule

\multirow{3}{*}{\parbox[c]{\hsize}{\emph{\underline{Maintenance}}}} &
*Deprecated & Presence of deprecated libraries that are no longer supported, lack documentation, or have been updated to newer versions
\\&
Technical-debt & Presence of quick fixes and temporary solutions accumulated over time
\\&
*Code-size & Overall size and complexity of the codebase
\\&
*Code-age & Age of the codebase
\\\bottomrule
\end{tabular}
\end{table*}

The context factors related to codebase characteristics are described in Table \ref{tab:codebase-factors}.

\subsubsection{Familiarity} 
Familiarity with a codebase—gained over time—fosters knowledge about its structure, common issues, and potential problem areas. This accumulated knowledge translates into faster and more efficient debugging. Familiarity with both the codebase and the underlying frameworks and libraries is crucial for effective troubleshooting. It allows developers to make informed decisions about debugging strategies, prioritize investigation areas, and ultimately resolve issues more quickly. This aligns with Gould's observation that developers take less time to debug when revisiting a codebase \cite{gould1975some}, and subsequent research highlighting the importance of program comprehension for successful debugging \cite{katz1987debugging, Ahmadzadeh2005AnAO,gugerty1986debugging, nanja1987analysis}.

Developers well acquainted with the codebase employ \textit{hypothesis-testing} efficiently (Figure \ref{fig:defect-factors-strategies}-d-YES). This expertise allows them to use techniques like \textit{binary-search} or \textit{simplification} to isolate irrelevant code and pinpoint problem areas. Understanding the interactions among different code elements, especially for asynchronous behavior, is key. One participant emphasized the importance of codebase familiarity for hypothesis generation: 
\begin{quote}
    \textit{``To form a precise hypothesis (...) you have to be very familiar with the code... how the software is put together (...) Working backwards from what the problem was and trying to come up with hypotheses that could explain how you got there.''}(P12)
\end{quote}

In contrast, when faced with an unfamiliar codebase, developers typically start with \textit{forward-reasoning} debugging (Figure \ref{fig:defect-factors-strategies}-d-k-NO) to understand its architecture, libraries and syntax. 
This foundational understanding helps to form initial hypotheses about the defect. However, the lack of prior knowledge hampers the formation of strong hypotheses, leading developers to use approaches like \textit{error-message} or \textit{backward-reasoning} debugging. Interestingly, our results showed that developers also explore tools like ChatGPT to gain insights into unfamiliar code, particularly for maintenance tasks or languages in which they are not specialists:
\begin{quote}
   \textit{``I may not even understand a particular language (...), I have started using ChatGPT where I would just put that piece of code, and it would help me understand what it does.''} (P6) 
\end{quote}
    
\subsubsection{Access}
Access to the codebase, the source code of the frameworks, the environments, and other resources significantly influences the choice of effective debugging strategies. 

\paragraph{Source code access:} 
The ability to view, edit and understand the source code, on a local machine or a server, is essential for effective debugging.
Debugging often relies heavily on direct access to the source code. When dealing with compiled binaries on production machines, this may not be available, limiting the ability to troubleshoot effectively. As one participant noted:
\begin{quote}
    \textit{``You would pretty much always do this (backward-reasoning) in a local machine (...) You probably cannot do that on production because it's gonna be all compiled binaries...''}(P8)
\end{quote}

\paragraph{Resources} Access to source code in different environments, such as development or production, affects the strategy selections. In addition, the level of access to these resources, as well as access to hardware and servers, plays a role in the way developers debug. Without access to the source code or detailed documentation, developers may struggle to understand how possible APIs function, making it difficult to pinpoint the cause of defects or implement fixes. Limited access to these resources can hinder the identification and resolution of defects, particularly if the API or library is no longer supported. The ability to replicate the exact environment locally or use emulators can be invaluable for isolating issues and using \textit{backward-reasoning}, \textit{binary-search or simplification} debugging.

\paragraph{Level of access} 
The level of access to the systems and components involved in the application, including the codebase, configurations, or deployment environments, significantly affects debugging strategies.
Full access allows for a comprehensive investigation, while partial or restricted access can limit the effectiveness of certain debugging approaches. This constraint requires developers to employ alternative strategies, such as indirect debugging methods or leveraging logs and metrics. For instance, developers with limited access may rely on logs and monitoring tools to infer the root cause of the problem.

\subsubsection{Maintenance}
Inadequate code maintenance presents significant challenges for developers, primarily due to deprecated code and the increased size and complexity of the codebase. Quick fixes and shortcuts implemented during development can have unforeseen consequences later, requiring additional debugging effort to resolve. Poorly maintained code often leads to significant technical debt, making debugging time consuming and diverting attention from new features or critical fixes.

\paragraph{Deprecated Code} 
When defects are related to deprecated code, developers prefer minimal modification techniques to avoid introducing new issues. They may use strategies like \textit{hypothesis-testing} and \textit{backward-reasoning} to understand and resolve the problem without altering the deprecated code itself. Some developers opt to \textit{replace} the deprecated code with recommended alternatives, aiming to both fix the defect and improve maintainability in the long run.

\paragraph{Code-size} 
Small codebases are easier to navigate and understand, even for developers unfamiliar with the codebase. This supports \textit{forward-reasoning} debugging, where the developer explores and understands the code structure before applying specific debugging techniques. 
Conversely, large and complex codebases require alternative strategies due to limited familiarity. In large and unfamiliar codebases, developers often rely on \textit{error-message} debugging, \textit{hypothesis-testing}, and \textit{simplification} techniques. In some cases, developers employ \textit{ code refactoring} to improve readability and structure without altering functionality. To isolate the problematic area, developers might generate log statements, though this can be overwhelming in highly complex codebases.

\subsection{Organizational Context}
Table \ref{tab:other-factors} summarizes the organizational factors that impact developers' debugging. As participants highlighted, the level of control over the environment available within an organization significantly affects their debugging choices. For example, debugging a third-party library or a service managed by an external team requires a different approach than when full access to the codebase is available.
\begin{quote}
    \textit{``Sometimes you're dealing with [a] third-party API (...) so you don't have control over that. And [so] you are limited [to] like binary search debugging and [to] remove stuff.''}(P10)
\end{quote}

Other factors beyond developer control, such as cost reduction and organizational incentive structures, influence the debugging behavior. When developers are primarily evaluated based on the number of issues they resolve, developers may feel pressure to debug in ways that sacrifice maintainability.

\subsection{Tool Availability and Usability}
The choice of debugging strategies is significantly influenced by the availability and type of tools used in the debugging process (Table \ref{tab:other-factors}). 
\textit{Version control systems}, through access to code commits and change history, allow developers to identify recent modifications and understand their potential impact on current issues. When available, \textit{logs} offer valuable insight into user interactions and system behavior, helping developers trace and analyze the flow of events leading to a defect. 

\begin{table*}[]
\centering
\sffamily
\fontsize{8}{8}\selectfont
\caption{\textbf{Organizational, Tools availability, Individual traits, and Project Requirements }factors influencing debugging approaches.
}

\label{tab:other-factors}
 \begin{tabular}{p{0.16\linewidth}p{0.78\linewidth}}
\toprule
\parbox[c]{\hsize}{\textbf{Factor}} & \textbf{Description}
\\\midrule
\multicolumn{2}{c}{\parbox[c]{\hsize}{\textbf{Organizational Context}}} 
\\\midrule
Manager-approval & Requirements for confirming changes with a manager or approver\cite{layman2013debugging}
\\
Developer-customer & Direct interaction between the development team and customers for requirements and feedback
\\
Control & Level of access and familiarity a developer has over code, libraries, team ownership, and other aspects of a project
\\
Support & Rewards based on the number of GitHub issues or customer tickets resolved; expert developers' availability 
\\
*Cost & Time, money, and resources required for troubleshooting\cite{evans2022pocket}
\\\toprule
\multicolumn{2}{c}{\parbox[c]{\hsize}{\textbf{Tools Availability and Usability}}}\\\midrule
Debugger & The applicability and difficulty of setting up the debugger\\

Logs & The availability of logs for monitoring user interactions and system behaviour
\\
*Version control & The availability of code commits and changes history \cite{evans2022pocket, SpinellisModernDebugging}
\\
Monitoring & 
The availability of monitoring tools and metrics, such as performance profilers, to track drastic changes 
\\
Online simulation & The availability of online simulation tools to replicate the error in specific environments, devices, and configurations
\\\toprule

\multicolumn{2}{c}{\parbox[c]{\hsize}{\textbf{Individual Developer Traits}}}\\\midrule

Experience & Developers expertise and Knowledge about the code, environment and language
\\
Habit &	Developers habitual characteristics that makes them feel more comfortable using one approach for all defects
\\\toprule

\multicolumn{2}{c}{\parbox[c]{\hsize}{\textbf{Project Requirements}}}

\\\midrule
Requirements & The availability of some restrictions like deadlines, and rules in work environment (e.g., agile development) or availability of documentation to guide debugging
\\
Aggregate& The availability of indicators of the normal behavior of a system
\\\bottomrule

\end{tabular}
\end{table*}

\subsection{Individual Developer Traits}
Developers’ characteristics, such as experience and habitual preferences, significantly impact their choice of debugging strategies. Our study showed that experienced developers' familiarity with various projects and issues allows them to make informed hypotheses and apply effective strategies. Some rely on structured approaches, while others use intuition and educated guesses based on their experience. Habits also play a role. One participant in our study favors certain debugging strategies based on past successes. 

\subsection{Project Requirements}
Participants in our study claimed that tight deadlines often force them to choose faster, less thorough debugging methods, which can increase the risk of overlooking underlying defects or introducing new ones. Additionally, the availability of project requirements influences how developers implement solutions and, consequently, their choice of debugging strategies. For instance, comprehensive documentation may lead to the creation of clear test cases that clarify expected system behavior and enable more effective debugging.

\section{Threats to validity}
Although our study provides a unique look at how expert developers consider contextual factors when choosing how to debug a problem, several limitations of our study must be considered when interpreting our results.

There were several issues in our study related to the validity of the construct. 
We collected programming problem scenarios from related works and participant reports in the first study. Since these resources are mainly for non-expert developers, the scenarios may not cover many problems tackled by expert developers in real programming situations.
Although we attempted to mitigate this bias by asking expert developers to describe their recent challenging debugging experiences, the list of problem scenarios, while representative of broad categories of challenging problems, encompassed specific tasks.  Future research should investigate explicit debugging strategies in a more comprehensive range of tasks. Conducting observational studies on expert developers working on their own challenging debugging tasks could also reveal different factors and strategies, providing deeper insight into the debugging process.

Some participants struggled to recall their last challenging debugging experience. This type of memory bias is inherent in most research methods. However, we attempted to control memory bias by asking interviewees to recall their most recent debugging experience and refer back to resources such as GitHub commit comments or chat history with team members.

From an external validity perspective, there were several issues. First, we attempted to recruit a diverse sample of developers to enhance generalizability. Our participant sample included developers with various backgrounds and levels of experience to ensure representation. This diversity introduced variation in the types of challenging problems faced, potentially diluting the focus on highly challenging problems encountered by expert developers. In our recruitment, we relied on the authors' social networks and snowball sampling. This might have led to sample bias, capturing unique demographics, and programming expertise not reflective of the broader developer population. There were no well-validated measures of prior programming knowledge or specific strategies, leading to coarse measures of developer expertise. As a result, our claims about developers' expertise are tentative. 

The second limitation is about the generalizability of the results beyond web development. Our study focused on web development, which is one of the most challenging environments. However, there are various other problem contexts that might have different characteristics that affect strategy choices. In addition, the list of debugging scenarios, while representative of broad categories of challenging problems, encompassed specific tasks. These may not fully reflect the wide range of issues in software engineering. Future research should investigate explicit debugging strategies in different domains and a more comprehensive range of tasks.


From an internal validity perspective, in our study we relied on self-reports for data collection, subjecting our findings to recall bias. To mitigate this, we asked follow-up questions and encouraged participants to refer to their programming environment and its history. Despite these efforts, memory bias may still have affected responses. In addition, utilizing a preliminary codebook might have influenced the responses of the participants. However, it was intended as a flexible guide rather than a rigid framework, allowing exploration of common themes without restricting the conversation. Standardized instructions and double-blind analysis were used to ensure that new themes and codes emerged independently.

\section{Discussion}
\label{sec:discussion}
Our study results expand our understanding of debugging, positioning it as an intricate, multifaceted process rather than a straightforward, fix-oriented task. 
Our observations with expert developers highlight how a wide range of factors about the defect itself, an individual developer, the team context, available tooling, and the codebase all come together to lead developers towards or away from specific debugging strategies. Our findings suggest that debugging is a deeply strategic process that is closely impacted by a wide range of factors, such as proactive code improvements, tests, and a deep understanding of the codebase. 

The results showed that when developers debug, they often employ multiple strategies, adapting their choices based on the information they gather about the context of the problem. Key context factors include familiarity with the codebase \cite{Ahmadzadeh2005AnAO, decasse1988review, gould1975some}, the availability of clear error messages \cite{spinellis2006debuggers, evans2022pocket, layman2013debugging, gould1975some}, and the testability of the system, including the availability of test cases \cite{layman2013debugging, bohme2017bug, SpinellisModernDebugging} that are consistent with previous research. In addition to these well-documented factors, our study reveals several other significant influences on strategy selection, including organizational considerations, individual traits, tool availability, and project requirements.

Organizational consideration involves costs, customer needs, manager approval, and the availability and type of support provided during the debugging process. These factors can significantly influence developers' strategy choices and can sometimes override individual preferences, such as habitual strategies or preferences for specific debugging approaches. Project requirements, tool availability, and the level of access a developer has over them often dictate the overall strategy to align with broader project goals and constraints.

In essence, while individual, organizational, tool availability and project requirements factors play a role in the selection of debugging strategies, the characteristics of the defect and the codebase predominantly impact how developers choose and adjust their debugging strategies, driving the dynamic adjustment and refinement of strategies as new information is uncovered during the debugging process.


Our study revealed that expert developers see debugging as an opportunity to enhance code quality through refactoring. Well-established test cases provide clear guidelines and help narrow down debugging efforts. This approach not only aids in selecting more appropriate debugging strategies, but also contributes to improved code maintainability. Thus, debugging becomes part of a broader quality improvement process than just a means of fixing isolated issues.

Although our results reveal some patterns in how contextual factors influence developers' decisions, developers may still consider other factors when debugging real-world problems. Future studies could evaluate these findings in a focus group setting while observing developers working on their own challenging problems in real time. Additionally, we discovered that developers often described debugging strategies that were not previously documented and found them to be more effective in certain situations. Sometimes, they referred to a strategy from our list, but mentioned that the strategy they would use is slightly different from the one we proposed. Future research should focus on enriching and expanding existing strategies to cover the scenarios mentioned by developers.

These insights have significant implications for software engineering researchers, educators, and practitioners. We discuss how our findings can be applied to these groups and outline opportunities for future work.


\textbf{For Educators.}
Our findings suggest that teaching debugging should extend beyond the simple use of debuggers and tools. Emphasis should be placed on enhancing students' decision-making skills by exploring the assumptions and prerequisites necessary to apply specific debugging approaches.

Educators should focus on training students to assess various contextual factors that could influence debugging strategies. This will allow students to make informed decisions when choosing the most suitable debugging approach. In addition, incorporating exercises that simulate real-world debugging scenarios and requiring students to identify and evaluate context factors before formulating a debugging strategy will enhance their cognitive skills and informed decision-making skills. 

The factors listed in Tables \ref{tab:defect-factors}-\ref{tab:other-factors} can serve as a valuable checklist for students to learn to choose appropriate strategies and refine their problem-solving skills. In addition, updating the curriculum to include modules on how organizational factors, such as cost and customer needs, influence debugging strategies will provide students with a holistic view of debugging within a larger project context.

\textbf{For practitioners.} Novice developers can take advantage of these insights to improve their debugging skills by applying the findings as a decision-making guide. This helps them better understand how to evaluate contextual factors and select the appropriate debugging strategies that, in the long term, significantly enhance their problem-solving abilities. 
More experienced engineers can extend this list with their own definitions and heuristics, potentially incorporating them into training materials for less experienced team members. This fosters continuous learning and cultivates a stronger debugging culture within the team.

\textbf{For researchers.} Our work opens avenues to explore the cognitive processes involved in debugging. Future research can benefit from our findings in different ways. First, it could investigate the thought processes that developers use when selecting debugging strategies. This could involve observational studies or cognitive interviews to understand how contextual factors are evaluated. Second, researchers can focus on developing and evaluating new debugging tools that assist developers in considering contextual factors when selecting strategies. These tools could provide real-time recommendations or automate parts of the decision-making process. Third, longitudinal studies to observe how developers' debugging strategies evolve over time with experience can be another venue to explore. This can provide insights into the learning curve associated with different debugging techniques and contextual evaluations.
This offers the opportunity to improve our understanding of debugging processes and improve training and support systems for developers at various stages of their careers.

\bibliographystyle{abbrv}
\bibliography{references}  






\end{document}